\documentclass{aastex}
\input psfig.sty
\shortauthors{F. Frontera et~al.}
\shorttitle{GRB990712 with BeppoSAX}

\newcommand{\sax}{{\it BeppoSAX}}
\newcommand{\ergcm}{\mbox{ erg cm$^{-2}$}}
\newcommand{\ergcms}{\mbox{erg cm$^{-2}$ s$^{-1}$}}
\def\etal{{\it et al. }}

\begin{document}

\title{The prompt  emission of GRB990712 with \sax: evidence
of a transient X--ray emission feature}

\author{F. Frontera\altaffilmark{1,2},
L.~Amati\altaffilmark{2},
M. Vietri \altaffilmark{3},
J.J.M.~in 't Zand\altaffilmark{4} 
E.~Costa\altaffilmark{5},
M.~Feroci\altaffilmark{3},
J.~Heise\altaffilmark{4},
N.~Masetti\altaffilmark{2},
L.~Nicastro\altaffilmark{6},
M.~Orlandini\altaffilmark{2},
E.~Palazzi\altaffilmark{2},
E.~Pian\altaffilmark{2},
L.~Piro\altaffilmark{5},
P.~Soffitta\altaffilmark{5}}

\altaffiltext{1}{Dipartimento di Fisica, Universit\`a di Ferrara, Via Paradiso
 12, 44100 Ferrara, Italy}

\altaffiltext{2}{Istituto Tecnologie e Studio Radiazioni Extraterrestri, 
CNR, Via Gobetti 101, 40129 Bologna, Italy}

\altaffiltext{3}{Dipartimento di Fisica, Terza Universit\`a di Roma,
via della Vasca Navale, 84, 00146 Roma}

\altaffiltext{4}{Space Research Organization in the Netherlands,
 Sorbonnelaan 2, 3584 CA Utrecht, The Netherlands}

\altaffiltext{5}{Istituto Astrofisica Spaziale, C.N.R., Via Fosso del 
Cavaliere, 00133 Roma, Italy}

\altaffiltext{6}{Istituto Fisica Cosmica e Applicazioni all'Informatica, 
C.N.R., Via U. La Malfa 153, 90146 Palermo, Italy}

\begin{abstract}
We report on the prompt X-- and $\gamma$--ray observations of 
GRB990712 with the \sax\ Gamma-Ray Burst Monitor and 
Wide Field Camera No. 2. Due to Sun constraints, we could not perform
a follow-up observation with the \sax\ Narrow Field Instruments. 
The light curve of the prompt emission shows two pulses and a total 
duration of 
about 40~s in X-rays. In gamma--rays the event is even shorter.
The 2--700 keV spectral emission with time shows a discontinuity in the peak 
energy
$E_p$ of the $E F(E)$ spectrum: $E_p$ is  above our energy passband  during 
the first pulse and goes down to $\sim 10$~keV during the second pulse.
Another peculiarity is noted in this event for the first time:
the evidence of a 2~s duration emission feature during the tail 
of the first pulse. The feature is consistent with either a Gaussian profile 
with centroid energy of 4.5~keV or a blackbody spectrum with $kT_{bb} 
\sim 1.3$~keV. We discuss the possible origin of the feature. The most
attractive possibility is that we are observing the thermal emission
of a baryon-loaded expanding fireball, when it becomes optically thin.

\end{abstract}


\section{Introduction}

Observations of cosmic Gamma-Ray Bursts (GRBs) with the \sax\
satellite are providing a key contribution to theories about their nature. 
Among the still unsettled questions, it is still not clear what are the mechanisms 
that 
produce the observed X--ray spectra and their evolution
with time (see, e.g., Frontera et al.\ 2000\nocite{Frontera00}) and what are the
environments in which the GRBs occur. In the context of the internal
shock model, synchrotron radiation is generally expected to play an important role in 
the
production of the observed GRB spectra (e.g., Tavani 1996), but Inverse Compton
can also give a significant contribution to them  \cite{Ghisellini00}. Also 
blackbody emission from the photosphere of the fireball 
\cite{Meszaros00} is expected to contribute to the GRB spectra, and inhomogeneities 
in the GRB outflow, made of  dense highly ionized metal--rich material,  
could give rise to broad spectral features, mainly K-edges \cite{Meszaros98}.
Effects of photoelectric absorption and Compton
scattering from the circumburst material can modify the intrinsic energy
spectrum of the GRBs, with the introduction of absorption cut-offs and
features, such as K-edges and emission lines
\cite{Meszaros98,Bottcher99}, the presence of which has already
been reported for some GRBs \cite{Yoshida99,Piro99,Antonelli00,Piro00,Amati00}.
The separation of the intrinsic and external components
is of key importance for establishing both the GRB emission mechanisms and
the properties of the GRB environment.

The Gamma Ray Burst Monitor (GRBM, 40--700~keV, Frontera et~al. 1997\nocite{Frontera97}), 
and the two Wide Field Cameras (WFC, 2--28~keV, Jager et~al.\ 1997\nocite{Jager97}) 
on board \sax\ offer the opportunity to study the GRB energy 
spectra in the 2--700~keV energy band where the above components can be
investigated (e.g., Frontera et al. 2000). 
GRB990712 was detected by the WFC No. 2 and the GRBM, showing in the 
2--26~keV band the highest peak flux ever observed from a GRB with \sax.
Its position was promptly distributed to the astronomical
community \cite{Frontera99}. Follow--up 
observations with the \sax\
Narrow Field Instruments were not possible  due to Sun
constraints. Observations  were performed in the optical and radio
bands. An optical transient (OT) with 
magnitude $R = 19.4\pm0.1$ was discovered about 3~hrs after the event
\cite{Bakos99} and its redshift is now well determined ($z = 0.4331\pm 0.0004$) 
\cite{Vreeswijk01}.
%

\section{Observations} 
\label{obs}
GRB990712 was detected  on July 12, 1999 
starting on 16:43:02 UT \cite{Frontera99}. Its position was determined with 
an error radius of $2'$ (99\% confidence level) and was centered at
$\alpha_{2000}\,=\, 22^{\rm h}31^{\rm m}50^{\rm s}$,
$\delta_{2000}\, =\, -73^\circ24'24''$ \cite{Heise99}.
Features and data available from GRBM and WFCs have already reported in several 
papers (e.g., Frontera et al.\ 2000 \nocite{Frontera00}).
The effective area exposed to the GRB was $\approx$~420~cm$^2$ in the
40--700~keV band and  37~cm$^2$ in the 2--26~keV energy band.
The background  in the WFC and GRBM energy bands  was fairly stable during
the event.
The WFC spectra were extracted through the Iterative Removal Of Sources
procedure (IROS, e.g. Jager et al. 1997 \nocite{Jager97}) which implicitly 
subtracts the contribution of the background and of other point sources in the 
field of view.
The count rate spectra  were  analyzed using the {\sc xspec} v.\ 10 software
package \cite{Arnaud96}. The quoted errors for the spectral parameters
correspond to 90\% confidence. Parameter
values shown in brackets in Table~1 have been fixed while fitting.

\section{Results}
\label{results}
Figure~1 shows the  time profile of GRB990712 in four energy
bands after the background subtraction. In all bands the GRB shows 
a double--pulse structure, with an opposite behavior with energy
of the first pulse with respect to the second: the peak flux of the first pulse 
increases with energy, while that of the second pulse decreases. 
The total duration of the event is about 20~s above 100~keV,  but much longer
(about 40~s) over the full 2--26~keV range. 
The spectral evolution of the event was studied by subdividing  the GRB time
profile into 8 adjacent time intervals and performing an analysis of the
spectrum of each interval (see Fig.~1). 
We fit the spectra  with either a power law ({\sc pl}, $N(E)\propto \ E^{\alpha}$),
a broken power--law ({\sc bkn--pl}) or a smoothly broken power law ({\sc bl}, 
Band et al.\ 1993\nocite{Band93}), with 
photoelectric absorption ({\sc wabs} model in XSPEC). We assumed the Galactic hydrogen 
column density $N_{\rm H}$ ($= 4.52\times 10^{20}\,$cm$^{-2}$, Dickey \& Lockman 1990 
\nocite{Dickey90}) along the line of sight to the GRB. For the first three time 
slices the fit was also performed leaving $N_{\rm H}$ free to vary, but
unfortunately $N_{\rm H}$  was not constrained by the data. The fit 
results with a {\sc pl} and a {\sc bl} are given in Table~1. The results  obtained 
with the {\sc bkn--pl} model were similar to those obtained with the {\sc bl} model and
are not reported in Table~1. 
The spectra of the time intervals A and B, that correspond to the rise of the 
first pulse of the GRB, and the spectrum of the interval D, that corresponds to the late
tail of this pulse, are well fitted with a {\sc pl} model. The {\sc bl} 
model provides a good  description of the spectra in the
time intervals E and F, that correspond to the core of the second pulse, while
the {\sc pl} model again well describes the tail of the second pulse (intervals G and
H). 
However neither of these models provides a good description of the C spectrum 
($\chi^2/\nu = 16/6$ for a {\sc pl} and 16/5 for a {\sc bl}). An excess with a 
significance level of about 3$\sigma$ is evident around 4 keV.
We investigated a possible instrumental origin of this anomaly with negative results. 
The WFC high-voltage, which is monitored every second, does not show any
glitch in the interval C. Given the accuracy of the readout (which dominates over
statistical noise) this excludes gain changes higher than 0.01\%. As in other GRB
detections, the WFC ratemeters all show the gamma-ray burst, even the ones that 
measure the illegal events, so we do not find any anomaly in this event.
There are no dips or spikes of any sort, with a typical 3$\sigma$ upper limit 
of 10\% per second of measurement (for a count rate of 700 counts s$^{-1}$).
There is one thermometer that shows a small change about 10~s before the
burst, but all other measurements do not show anything out of the ordinary.
The count excess in the GRB spectrum of the interval C is also clearly
 visible (see Figure 2) in the ratio between 
the C count spectrum and the Crab spectrum measured  when this source was observed 
at an angular offset similar to that of GRB990712.  For comparison, Figure 2 also 
shows 
the ratio with Crab of the spectra measured in the intervals B and D, that precede
and follow C, respectively. As can be seen, in the interval B the greater
hardness of the GRB spectrum (photon index of $\sim$1.4, see Table 1) than that
of Crab is apparent,
while in the interval D  the flatness of the Crab ratio is consistent with the
similar slope of the GRB spectrum (see Table~1) with that of Crab. A slight hint
of the 4~keV excess in the spectrum also  appears in the interval B, but it is not 
statistically significant.  
We point out that the Crab ratio technique is  adopted, 
to discover cyclotron lines in the spectra of X-ray pulsars (e.g., Dal Fiume
et al. 2000\nocite{Dalfiume00}). 

The addition of a Gaussian function or a blackbody spectrum ({\sc bb}) to a {\sc pl} 
model 
provides a good fit ($\chi^2/\nu = 1.6/5$ and $\chi^2/\nu = 6.2/6$,
respectively) of the C spectrum. The best fit parameters of both
models are reported in Table~2. For a better determination of the Gaussian and 
{\sc bb}
parameters, the photon index $\alpha$ of the {\sc pl} model was kept fixed in the fit to
the best fit value, that is given by
$1.34 \pm 0.17$ or $1.24 \pm 0.20$, depending on the model assumed
for the feature, a Gaussian or a {\sc bb} model, respectively. 
The count rate spectrum of the interval C along with the best fit curve of the 
{\sc pl} plus
a blackbody model is shown in the top panel of the Figure 3, while the ratio
between the count spectrum and the best fit power-law alone is shown in the bottom 
panel. The excess counts to the {\sc pl} model are apparent.
The evolution of the logarithmic power per photon decade (the $E F(E)$ spectrum) 
with the time from the GRB onset is shown in 
Figure~3. The emission feature during the C interval is also apparent in this plot.
The peak energy of the $E F(E)$ spectrum (see Table 1) is above 
our energy passband for the entire duration of the first pulse before
suddenly becoming  much lower ($\sim 10$ keV) from the beginning of the second pulse. 

From the spectral fits we derived GRB fluence and peak flux.
The $\gamma$--ray (40--700~keV) fluence of the burst is $S_\gamma\,=\,
(6.5 \pm 0.3)\times 10^{-6}$ \ergcm, while the corresponding value found
in the 2--10 keV band is $S_{\rm X} = (2.60 \pm 0.06)\times 10^{-6}$~\ergcm, with
a ratio $S_{\rm X}/S_\gamma = 0.40\pm 0.03$, which is one of the highest values 
found with
\sax\/ \cite{Frontera00}. The 2--700~keV fluence is given 
by $S_\gamma\,=\, (1.10 \pm 0.03) \times 10^{-5}$~\ergcm.
The $\gamma$--ray peak flux is $P_\gamma = 4.1\pm 0.3$~photons/cm$^2$~s
corresponding to $(1.3 \pm 0.1)\times 10^{-6}$~\ergcms, while the
corresponding 2--10 keV peak flux is  $P_{\rm X} = 41\pm4 $~photons/cm$^2$~s,
corresponding to $(3.3\pm 0.3)\times 10^{-7}$~\ergcms.

\section{Discussion}
\label{disc}
%
%
%
From the redshift value of the optical afterglow of GRB990712 ($z = 0.4331$,
Vreeswijk et al.\ 2001\nocite{Vreeswijk01}) we can derive the X-- plus
$\gamma$--ray energy released in the main event. Assuming isotropic
emission and a standard Friedman cosmology ($H_0 = 70$~km~s$^{-1}$
Mpc$^{-1}$ and $q_0 = 0.5$), we get a 2--700~keV released energy 
of $E_{rel} = (5.9 \pm 0.2 ) \times
10^{51}$~erg. A sizeable fraction of this  energy ($\sim$20\%)
is released between 2 and 10 keV. If we exclude the controversial case
of GRB980425/SN1998bw \cite{Galama98,Kulkarni98,Pian00}, GRB990712, 
in addition to showing the lowest redshift, is  one of the least energetic events. 

GRB990712 is marked by a peculiarity, which is noted here for the
first time: the evidence of a broad emission feature
at 4.4~keV, visible for 2~s, superposed on a power law continuum model. 
It can be described
by either a Gaussian profile with a full width at half maximum of
$\sim$3~keV or a blackbody emission with $kT_{bb} = 1.3$~keV. 
If we assume the Gaussian description, its centroid energy
corrected for redshift ($E_0 = 6.4\pm 1.1$~keV) is consistent
with the energy of both an iron K fluorescence line and a iron recombination line. 
The interpretation of the emission feature as an iron recombination line is tempting, 
yet it makes some stringent demands on models: in fact, it requires much mass to
be present within a few light seconds of the burst site, and for 
this mass to be moving at Newtonian speeds. 
Vietri et al (2001)\nocite{Vietri01} derive the expected rate of photons for a narrow
line:
\begin{equation}
\label{main}
\dot{N}_{Fe} \approx \frac{4 \times 10^{52}}{T_7^{3/4}}\; s^{-1}\frac{M_{Fe}}{1 M_\odot}
\frac{6\times 10^{15}\; cm}{R}
\end{equation}
where $T_7$ and $R$ are the electron temperature (units of $10^7$~K) and the 
external radius of the line
emitting medium, respectively, and $M_{\rm Fe}$ is the iron mass present in this medium.
For a broad line, such as that in GRB990712, the
above value is an understimate by a factor of a few at most. Assuming $T_7=1$, 
$R= 6\times 10^{15}$~cm and $M_{\rm Fe}= 1 M_\odot$,
comparison
with Table 2 shows that Eq. \ref{main} underestimates the observed line luminosity 
by four to five orders of 
magnitude; inserting $R = 10$ light seconds in the above equation yields the correct 
line luminosity, but for a total mass of iron of {\it at least}
$M_{Fe} = 0.1 M_\odot$. Assuming a realistic iron relative abundance
(at least 1\% of the total mass, if what we are seeing is a type I SN, but
more for any other hypothesis) shows that we must explain the presence 
of at least $10 M_\odot$ of matter, at radii of a few light seconds,
with none of this obstructing the line of sight. 

A more attractive possibility is that the observed feature is 
indeed thermal. Though we cannot definitely establish the thermal 
nature of the emission during the C interval, 
we wish to remark that the fireball model can account naturally
for the presence of these features. In fact, as remarked by 
both Paczynski (1986)\nocite{Paczynski86} and Goodman (1986)\nocite{Goodman86} 
hyper--relativistic 
expansion naturally leads to the liberation of a fully thermal 
spectrum, at the time when the fireball expansion becomes optically 
thin. The initial absence,
and later disappearance, of the peak in question does not create difficulties
within the fireball model: it can easily be ascribed to inhomogeneities
in the time--structure of the relativistic wind, inhomogeneities
which are in any case required in order to account for the burst
sub--second variability. The fact that the peak, furthermore, appears
during the tail of the first pulse of course makes the detection of
the weaker thermal component easier (see the very revealing Fig. 2 of
M${\rm \grave{e}}$sz${\rm \grave{a}}$ros and Rees 2000). 
If this spectral feature is indeed thermal in origin, within the
fireball model there is quantitative, and independent check on 
this identification. In fact, the expected photospheric radius within
the fireball model is \cite{Meszaros00}
$r_{ph} = 1.2\times 10^{13}\; cm\; L_{52} Y  \eta_2^{-3}$,
where $L_{52}$ is the wind luminosity in units of $10^{52}$ erg s$^{-1}$,
$Y\approx 1$ is the number of electrons per baryon, and $\eta_2$ is the
flow Lorentz factor $\eta$ in units of $100$.
 At this radius, the observed photospheric luminosity and temperature 
are $L_{ph} = L_{52} (\eta/\eta_\star)^{8/3}$
and $\Theta_{ph} = \Theta_0 (\eta/\eta_\star)^{8/3}$, respectively,
where $\eta_\star \approx 10^3 (L_{52} \mu_1^{-1} Y \Gamma_0)^{1/4}$ and $\Theta_0 = 
1236 (L_{52}^{1/2} \mu_1^{-1} \Gamma_0^{-1})^{1/2}$~keV, $\Gamma_0$ ($\ge 1$) being the 
initial bulk Lorentz factor of the wind, and $\mu_1$ the mass, in units of 10~$M_\odot$, 
of the rotating black hole, from 6 times the gravitational radius of which the 
fireball is assumed to start its expansion. 
By fitting simultaneously the  observed temperature corrected for the redshift 
($1.86$~keV) and the luminosity of the photosphere ($\sim 2 \times 
10^{50}$ erg s$^{-1}$), we find $\eta \approx 100 Y^{1/4} (\mu_1^{-1} 
\Gamma_0^{11})^{1/54}$ and $L_{52} \approx 2 (\mu_1 \Gamma_0^{11})^{2/9}$.
We thus see that the two independently determined observational parameters,
blackbody temperature and luminosity, are well--fitted by values of
the theoretical parameters, $\eta$ and $L_{52}$, well within the 
expected range.  Assuming unit values for $Y$, $\mu_1$ and $\Gamma_0$, $L_{52}$ results 
to be about 100 times higher than the estimated 2--700~keV luminosity 
($\sim 2 \times 10^{50}$~erg s$^{-1}$) assuming isotropy. That would imply an 
efficiency of  only 1\% in the production of electromagnetic radiation.

In addition to the transient feature, the event shows a spectral
evolution that is not typical of other GRBs observed with \sax\/ \cite{Frontera00}:
the peak energy $E_p$ of the $E F(E)$ spectrum (see Table~1 and Figure~3) is 
constantly 
above our energy passband for the entire duration of the first pulse, while it
takes a low value ($\sim 10$~keV) with the onset of the second pulse.
This discontinuity  can be the result of two successive
electron acceleration episodes, giving rise to the first and the second pulses. 
The different 
peak flux behavior of the two pulses with energy, discussed in section 3 confirms 
this scenario.
The emission feature is found only during the first acceleration event.

\acknowledgements
Thanks to John Stephen for his careful reading of the manuscript. Also many
thanks to the anonymous referee who greatly stimulated us to improve 
the paper.

\clearpage

\figcaption[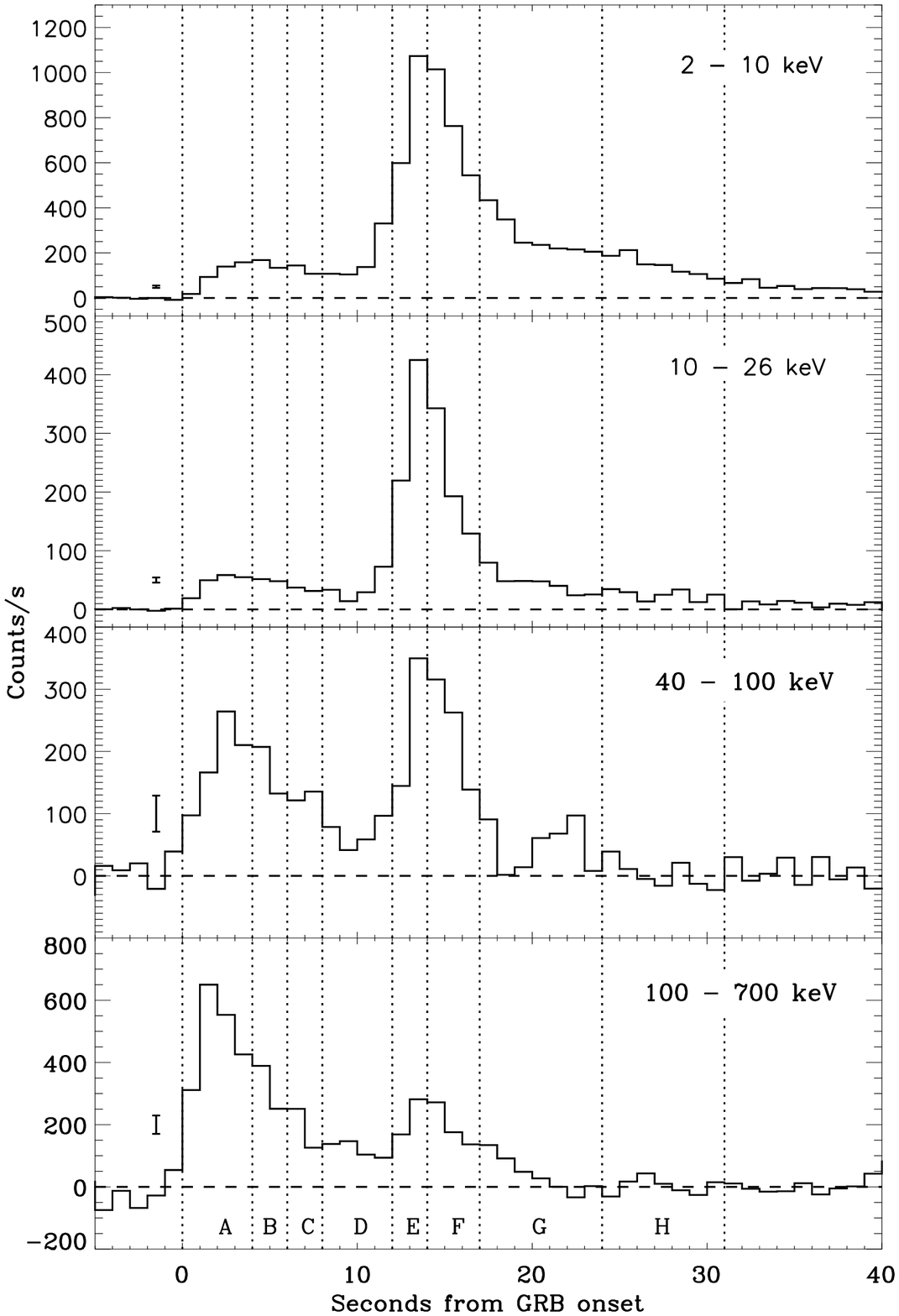]{Light curves of GRB990712 in four energy bands,
after background subtraction. The zero abscissa corresponds to 1999 July 12,
16:43:01.6 UT. The time intervals over which the spectral 
analysis was performed are indicated with vertical dotted lines.
\label{fig1}}

\figcaption[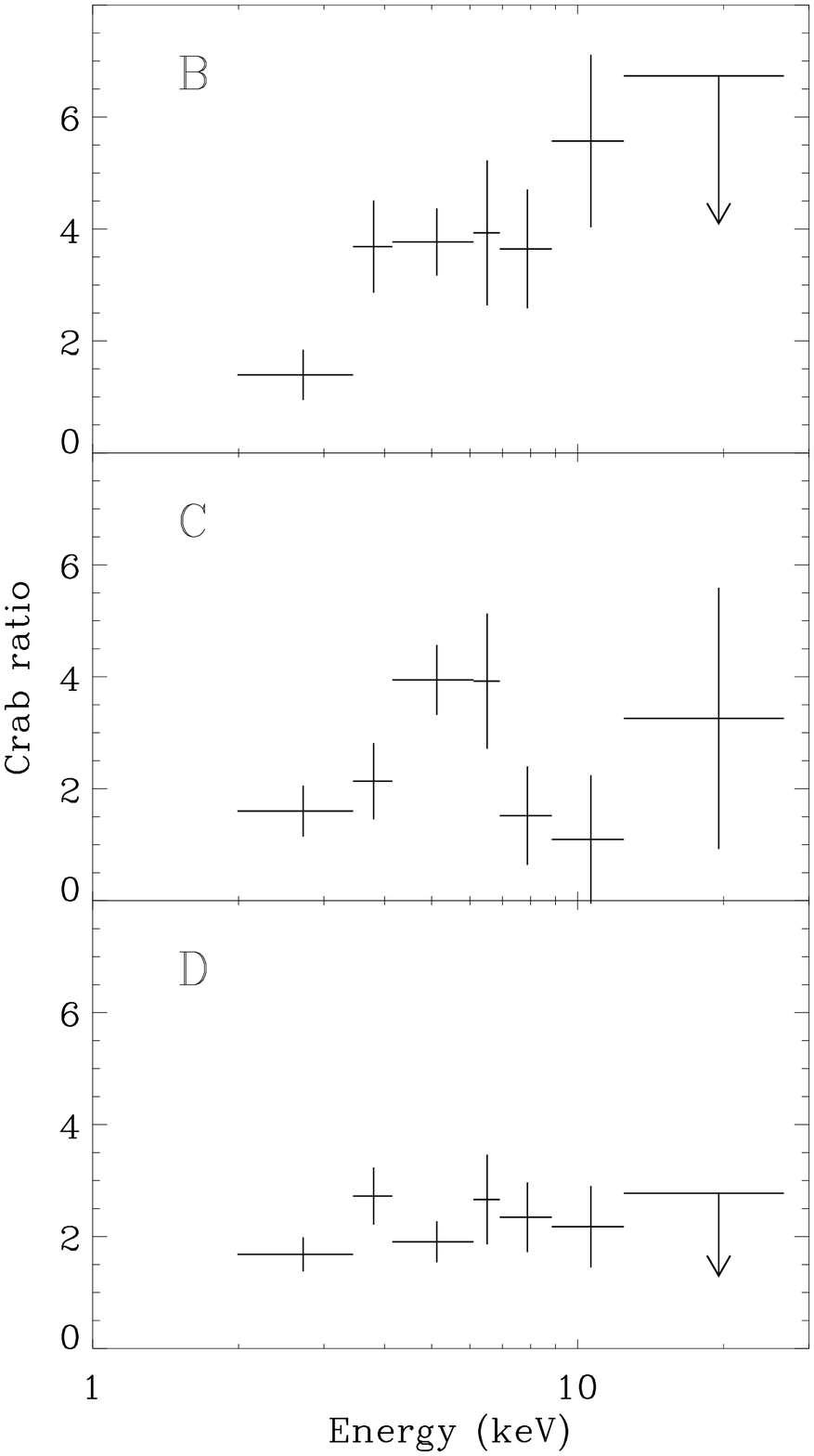]{Ratio with the Crab count spectrum of the GRB990712 spectra 
in the time intervals B, C and D, respectively, measured with the WFC No. 2. 
The Crab spectrum used was measured
when this source was observed  at an angular offset similar to that of GRB990712.  
\label{fig2}}

\figcaption[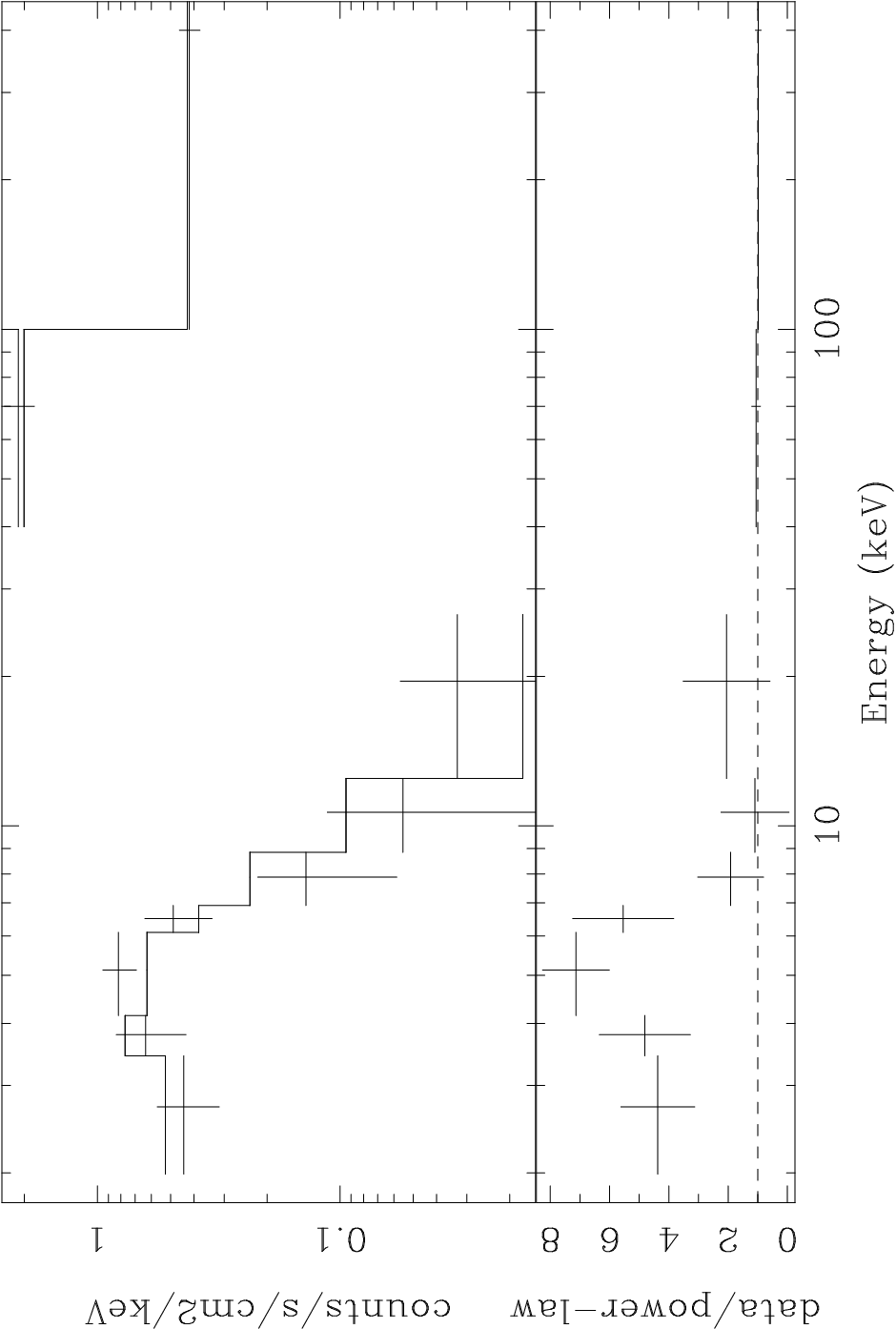]{ WFC + GRBM spectrum of GRB990712 in the time interval C, 
along with
the best fit curve obtained assuming a power--law plus a blackbody model. 
{\it Bottom panel}: 
ratio between the data and the best fit power--law model alone. \label{fig3}}

\figcaption[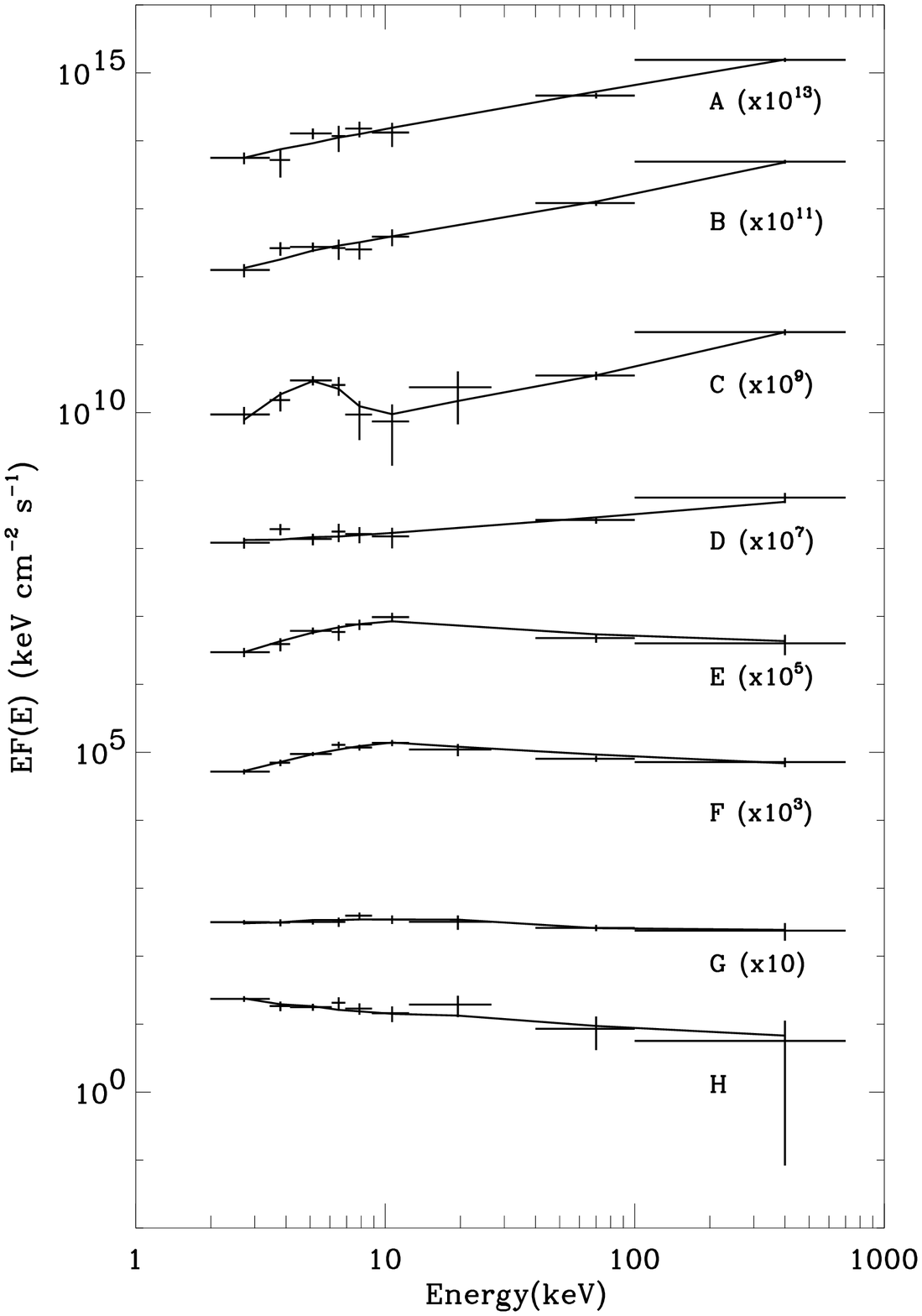]{$EF(E)$ spectrum of GRB990712 in the time intervals in which
we divided the burst time profile (see also Table~1). It is apparent the GRB 
peculiar spectrum in the interval C. \label{fig4}}

\clearpage

%
\begin{deluxetable}{cccccccc}
\tablewidth{0pt}
\tablenum{1}
\tablecaption{Spectral evolution of GRB990712 prompt emission}
\label{table1}
\tablehead{
Slice & Duration (s) & Model\tablenotemark{(a)} & $N_{\rm H}$\tablenotemark{(b)}
& $\alpha$ & $\beta$ & E$_{p}$(keV) & $\chi^{2}/\nu$
} 
\startdata
 A     & 4  &{\sc pl}   & 3.2$\pm$2.5 & $-1.40 \pm 0.09$  &  & $>700$ & 7.8/6   \\
       &    & {\sc pl}  & [0.0452]    & $-1.34 \pm 0.07$  &  &  & 10./7   \\

 B     & 2  &{\sc pl}   & 2.7$\pm$2.1 & $-1.44 \pm 0.08$  &  &$>700$  & 4.4/5   \\
       &    & {\sc pl}  & [0.0452]    & $-1.38 \pm 0.06$  &  &  & 7/6    \\
 C     & 2  &{\sc pl}   & 0.3$\pm$2.0 & $-1.66 \pm 0.11$  &  &$>700$  & 16./6   \\
       &    & {\sc pl}  & [0.0452]    & $-1.64 \pm 0.07$  &  &  & 16/7  \\
 D     & 4  &  {\sc pl} & [0.0452]    & $-1.80 \pm 0.07$  &  &$>700$  & 4.6/6  \\
 E     & 2  &   {\sc pl}& [0.0452]    & $-2.04 \pm 0.05$  &  &  & 37/7   \\
       & 2  &{\sc bl}   &[0.0452]     & $-0.4 \pm 0.7  $  & $-2.4\pm  0.3$   &
     11$\pm$8    & 3.5/5  \\
 F     & 3  &{\sc pl}   &[0.0452]     & $-2.01 \pm 0.04$  &  &  & 89/7   \\ 
       & 3 & {\sc bl}   &[0.0452]     & $-0.7 \pm 0.4$    & $-2.3 \pm 0.2$  &
      7$\pm$3    & 4.3/5  \\
 G     & 7 & {\sc pl}   & [0.0452]    & $-2.15 \pm 0.05$  &  &  & 11/7   \\ 
        & 7 &  {\sc bl} & [0.0452]    & $-1.6 \pm 0.6$    & $-2.3 \pm 0.3$  &
      20$\pm$15    & 2.5/5  \\
 H      & 7 & {\sc pl}  & [0.0452]    & $-2.3 \pm 0.2$    &  &  & 3.0/7  \\
\enddata
\tablenotetext{(a)}{The BL (Band Law) refers to 
the smoothed broken power-law proposed by Band et al. (1993): $\alpha$
and $\beta$ are the power--law photon 
indices below and above the break energy E$_{0}$, respectively, and
E$_{p}$=E$_{0}$(2+$\alpha$) is the peak energy of the $E F(E)$ spectrum.}
\tablenotetext{(b)}{$N_{\rm H}$ values are given in units of $10^{22}$~cm$^
{-2}$.}
\end{deluxetable}

\clearpage

%
%
\begin{deluxetable}{l c c}
\tablenum{2}
\tablewidth{0pt}
\tablecaption{Best fit parameters of the time slice C spectrum}
\tablehead{
\colhead{Parameter}    & \colhead{{\sc PL}+Gaussian} &  \colhead{{\sc PL+BB}
} \\
}
\startdata
$\alpha$                         &  [1.34]      & [1.24]                 \\
$F_{\rm PL} (@ 1\,{\rm keV})$ [cm$^{-2}$ s$^{-1}$] & $1.95 \pm 0.24 $ &
$1.13 \pm 0.14$    \\
$E_{\rm line}$ [keV]                & $4.4 \pm 0.8$         &                \\
$\sigma_{\rm line}$ [keV]           & $1.4 \pm 0.7$         &                \\
$I_{\rm line}$ [$10^{-8}$~erg cm$^{-2}$ s$^{-1}$] & $2.7 \pm 1.1$ &           \\
$L_{\rm line}$ [10$^{57}$ phot s$^{-1}$] & $2.5 \pm 0.9$         &           \\
$kT_{\rm bb}$ [keV]                 &                       &  $1.3\pm 0.3$   \\
L$_{bb}$ [10$^{49}$ erg s$^{-1}$] &                      &  $2.5 \pm 0.6$  \\  
$\chi ^2/\nu$                    & 1.6/5                 &  6.2/6          \\
\enddata
\end{deluxetable}

\clearpage

\begin{figure}
\centerline{\psfig{figure=fig1.ps,width=12.5cm,angle=0}}
\vspace{2cm}
\centerline{Fig. 1}
\end{figure}

\clearpage

\begin{figure}
\centerline{\psfig{figure=fig2.ps,width=12.5cm,angle=0}}
\vspace{2cm}
\centerline{Fig. 2}
\end{figure}

\clearpage

\begin{figure}
\centerline{\psfig{figure=fig3.ps,width=14.5cm,angle=-90}}
\vspace{2cm}
\centerline{Fig. 3}
\end{figure}

\clearpage

\begin{figure}
\centerline{\psfig{figure=fig4.ps,width=14.5cm,angle=0}}
\vspace{2cm}
\centerline{Fig. 4}
\end{figure}

\end{document}